\documentclass[aip,jcp,reprint,showpacs,singlecolumn]{revtex4-1}
\usepackage[T1]{fontenc}       

\usepackage[T1]{fontenc}
\usepackage[font=small,labelfont=bf,tableposition=top]{caption}
\usepackage{dcolumn}
\usepackage{placeins}
\usepackage{epsf}
\usepackage{subfigure}
\usepackage{epsfig,amsopn}
\usepackage{amssymb,graphicx}
\usepackage{amsmath,amsfonts}
\usepackage{graphicx}
\usepackage{dcolumn}
\usepackage{bm}
\usepackage{ifpdf}
\usepackage{natbib}
\usepackage{wasysym}
\usepackage[byname]{smartref}
\usepackage{color}
\usepackage{lipsum}
\usepackage{nicefrac}
\usepackage[breaklinks, colorlinks, linkcolor=blue, citecolor=blue, urlcolor=blue]{hyperref}
\usepackage[table]{xcolor}
\usepackage{multirow}
\usepackage{braket}
\usepackage{float}
\usepackage{comment}
\usepackage{verbatim}
\usepackage{enumerate}
\usepackage{enumerate}
\usepackage{tabularx}
\usepackage[table]{xcolor}
\DeclareCaptionLabelFormat{andtable}{#1~#2  \&  \tablename~\thetable}

\renewcommand{\v}[1]{\ensuremath{\mathbf{#1}}} 

\newcommand{\beq}{\begin{equation}}
\newcommand{\eeq}{\end{equation}}

\newcommand{\bea}{\begin{eqnarray}}
\newcommand{\eea}{\end{eqnarray}}

\begin{document} 

\author{Roman Korol and Dvira Segal}
\affiliation{
Department of Chemistry and Centre for Quantum Information and Quantum Control,
University of Toronto, 80 Saint George St., Toronto, Ontario, Canada M5S 3H6
}

\email{dvira.segal@utoronto.ca}

\title{Machine Learning Prediction of DNA Charge Transport}



\date{\today}
	
\begin{abstract}
First principle calculations of charge transfer in DNA molecules 
are computationally expensive given that charge carriers migrate in interaction with intra- and inter-molecular atomic motion. Screening sequences, e.g. to identify excellent electrical conductors
is challenging even when adopting coarse-grained models and 
effective computational schemes that do not explicitly describe atomic dynamics.
In this work, we present a machine learning (ML) model that allows  
the inexpensive prediction of the electrical conductance of millions of 
{\it long} double-stranded DNA (dsDNA) sequences, reducing computational costs by orders of magnitude.
The algorithm is trained on {\it short} DNA nanojunctions
with $n=3-7$ base pairs. The electrical conductance of the training set is 
computed with a quantum scattering method, which captures charge-nuclei scattering processes. 
We demonstrate that the ML method accurately predicts the electrical conductance of 
varied dsDNA junctions tracing different transport mechanisms: coherent (short-range) 
quantum tunneling, on-resonance (ballistic) transport, and incoherent site-to-site hopping.
Furthermore, the ML approach supports physical observations  
that clusters of nucleotides regulate DNA transport behavior. 
The input features tested in this work could be used in 
other ML studies of charge transport in complex polymers,
in the search for promising electronic and thermoelectric materials.
\end{abstract}

\maketitle
	
\section{Introduction}
\label{SecI}
	

Charge migration in DNA is a topic of significant interest with applications to biology, chemistry, 
physics and engineering \cite{Tapash}. First and foremost, both oxidative DNA damage, 
leading to cancerous mutations \cite{Bartonrev10,BartonRev}, and repair signaling \cite{Barton18} 
take place through long-range electron transfer.
Beyond its biological function, DNA is an attractive material in nanotechnology 
given its molecular recognition properties, self-assembly, and structural flexibility \cite{bookDNA}.
For example, self assembled monolayers of DNA can serve as biosensors for detecting mutated 
genes \cite{Bartonsensor}, and DNA templates can direct the assembly of nanostructures of desired shapes \cite{revDNA,bookDNA}. 
Moreover, given its rich electronic properties and automated synthesis, DNA molecules may be used as 
the conducting material in molecular electronic circuits \cite{BeratanRev,Ron,Danny}. 

Experiments have demonstrated diverse charge transport behavior through DNA.
Very long DNA molecules with hundreds of base-pairs (bp) may act like
metals, semiconductors, and even insulators \cite{Danny}.
Measurements of shorter (5-20 bp) DNA duplexes 
revealed a broad range of trends: Depending on the base sequence \cite{Tao04,Arianna,Lewis08,Lewis},  
molecular length \cite{Kelly,Giese,Tao04,Slinker, Tao16},
backbone composition \cite{Waldeck17}, surrounding environment \cite{Shao16}, 
temperature, helical conformation \cite{Hihath15}
linkers to the electrodes \cite{Harrishole, Waldeck17,Wagner17} and 
gating \cite{Taogate}, DNA charge transport may occur via different mechanisms:
deep tunneling  \cite{Tao04,Tao16}, thermally activated hopping \cite{Tao04,Tao16,Slinker},
resonant ballistic or flickering resonance \cite{flickering,BartonNN11},
and intermediate coherent-incoherent behavior \cite{Berlininter,Ratner15,Beratan16,HihathA}.

%

Calculations of DNA charge transfer are exceptionally challenging given the complexity of the system,
 with charge carriers interacting with moving nuclei of the base pairs, backbone, counterions, solvent.
Screening DNA sequences to identify excellent or poor conductors  and classify transport mechanisms
is challenging even when adopting coarse-grained models and computational schemes that only approximately
describe atomic dynamics. 
The elementary components of DNA are four nucleotides forming double-stranded helix DNA (dsDNA). 
The strands are hybridized by obeying the base-pairing rules: adenine (A) with thymine (T),
cytosine (C) with guanine (G).
Since the sequence space grows exponentially fast, brute-force all-sequence calculations of charge transport in DNA
become essentially implausible for chains with $n\gtrsim 8$ base pairs.

\begin{figure*}[htbp]
\includegraphics[width=15cm]{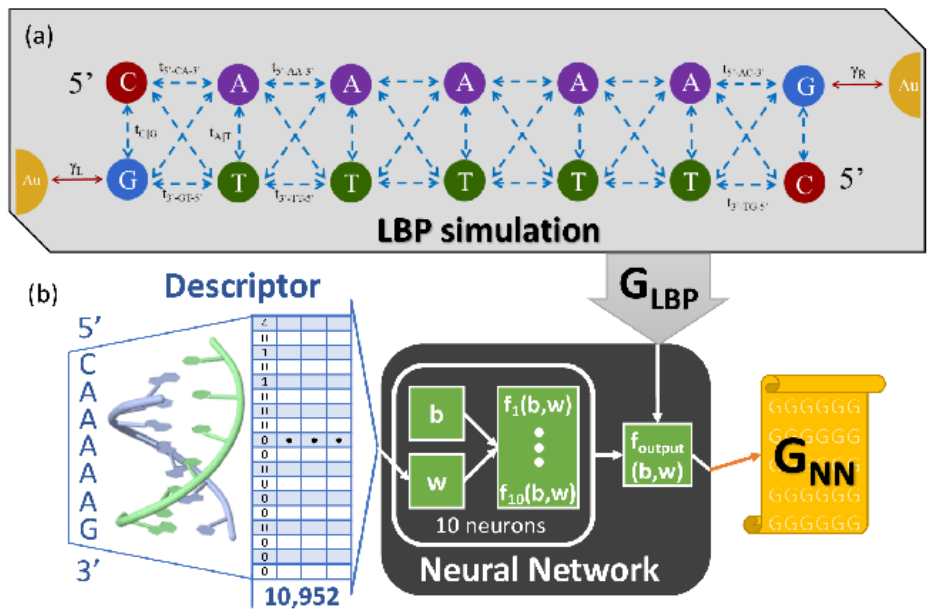}
\vspace{-10mm}
\caption{Machine learning calculation of the electrical conductance of dsDNA junctions.
(a) A coarse grained model of a dsDNA junction, depicting the $n=7$ bp sequence 5'CAAAAAG3'. 
Sequences are labeled from the 5' to the 3' end.
The electronic Hamiltonian is described with a tight binding model; 
double arrows represent electronic tunneling elements within and in between strands. 
The effect of the nuclear degrees of freedom is taken into account using a quantum
scattering approach, the LBP method. The output of an LBP calculation is the electrical conductance $G_{LBP}$.
(b) The neural network is trained based on the conductances of 10,592 sequences $n=3-7$ bp long.
The descriptor exemplified here has 16+2 entries: It counts the occurrence of each of the 16 possible pairs
of  nucleotides in a single strand in alphabetical order.
The sequence 5'CAAAAAG3' depicted here has four pairs of AA (first entry), a single AG pair (third entry),
and a single CA pair (fifth entry).
The last two entries of the input (0 here) identify the base that is connected to the metals, G,A,T or C,
taking the values 0,1,2 or 3, respectively.
}
\label{Fig1}
\end{figure*}

The objective of the present work is to test a machine learning (ML) model 
for DNA charge transport so as to bypass computational limitations of direct 
comprehensive calculations. The model predicts the conductance of millions of long 
metal-molecule-metal DNA junctions
based on conductance data of thousands of short (3-7 bp) sequences.
Our goal is to use the ML method to provide reasonable numbers for DNA electrical conductance and furthermore
predict correct qualitative transport trends for special classes of molecules---in accord with physical considerations.
For example, short DNA with an AT block should act as a tunneling barrier with the conductance 
dropping exponentially with length. In contrast, 
GC-rich molecules are expected to behave as ohmic conductors with the resistance growing linearly with the number of base pairs 
\cite{Tao16}.
Conductance values in DNA molecules extend over many orders of magnitude, with metallic to insulating sequences.
The ML model should capture this extreme variability.

The physical setup under consideration is depicted in Fig. \ref{Fig1}:
A dsDNA molecule with $n$ base pairs is connected to metal electrodes at its 3' ends.
Assuming molecules are coupled symmetrically to two identical 
metals, there are 8192 distinct junctions for e.g. molecules with $n=7$ bp \cite{Korol3}.
The conductance of these junctions (as well as of shorter systems) 
was computed and analyzed in our recent study, Ref. \cite{Korol3} 
based on a quantum scattering technique, the Landauer-B\"uttiker's probe (LBP)
method \cite{Buttiker1,Buttiker2}. 
Here we use this dataset to train a ML model so as to compute the conductance of arbitrarily 
long DNA junctions.
To make the model transferable to sequences of different size,
we prepare input features (descriptors) that do not rightly announce on the base sequence itself,
but report on molecular composition and the occurrence of local clusters. 


We train the ML model on the conductance of 3 to 7 bp sequences. We test it 
against short molecules in this set, then use it for predicting the  
behavior of longer molecules up to 18 bp. 
The quality of predictions is assessed against LBP calculations.
Our finding are that the ML model correctly predicts both global features of the ensemble of molecules,
and the behavior of special sequences.
Specifically, the model traces the transition from fully coherent (deep tunneling) to ballistic (on-resonance) transfer 
for stacked A-sequences, the tunneling to hopping crossover in sequences with an 
AT block, and (with mixed success) hopping conduction in alternating GC-sequences.
These results demonstrate that ML models can provide a cheap assessment 
of quantum dynamics and transport in complex systems.

%
%
	
The paper is organized as follows. 
In Sec. \ref{Method} we discuss the Landauer-B\"uttiker probe method 
and present the machine learning framework. 
In Sec. \ref{Result} we illustrate the performance of our neural networks for short sequences, 
as well as the extrapolation onto longer sequences. 
We further demonstrate that when used appropriately, 
the neural networks can learn the mechanism of conduction and accurately predict
the conductance of molecules that are significantly longer than sequences in the training set.
We conclude in Sec. \ref{Summ}.
	
\section{Methods}
\label{Method}

\subsection{Physical setup}

Loosely speaking, we classify DNA charge transport as `short' or `long' if it extends at most 2 nm ($n<7$ bp), 
or up to 10 nm ($7<n<30$ bp), respectively; recall that the distance between base pairs of DNA 
is approximately $3.4$ Angstroms.
Ultra long molecules extend beyond 10 nm, and are not examined in this study, 
though our approach is applicable to arbitrary length. 
Single-molecule conductance experiments in a metal-molecule-metal geometry
typically employ DNA molecules in the range of $7<n<20$ \cite{Tao04,Tao16,Waldeck17,Hihath15,Taogate,Ratner15}.
In contrast, DNA electrochemistry experiments on monolayers probe transport in 
molecules with as many as 100 bp, see e.g. Ref. \cite{BartonNN11}.

The prototype molecular junction under consideration is depicted in Fig. \ref{Fig1}(a).
A single dsDNA molecule with $n$ base pairs is connected to metal electrodes at its 3' ends.
While we sketch the DNA as if it directly attaches to the metals,
in fact in e.g. break junction experiments the molecule is modified with linkers (thiols, amines)
to ensure sufficiently strong chemical binding of the molecule to the electrodes.
The overall-effective coupling energy of the nucleotide at the interface to the metal is captured by
the hybridization parameters $\gamma_{L}$ and $\gamma_R$.
Nuclear degrees of freedom are not pictured in Fig. \ref{Fig1}. 
Implementing their effect on charge conduction is explained in the next section.


\begin{figure*}[htbp]
\vspace{-2mm} \hspace{-7mm}
\includegraphics[width=11cm]{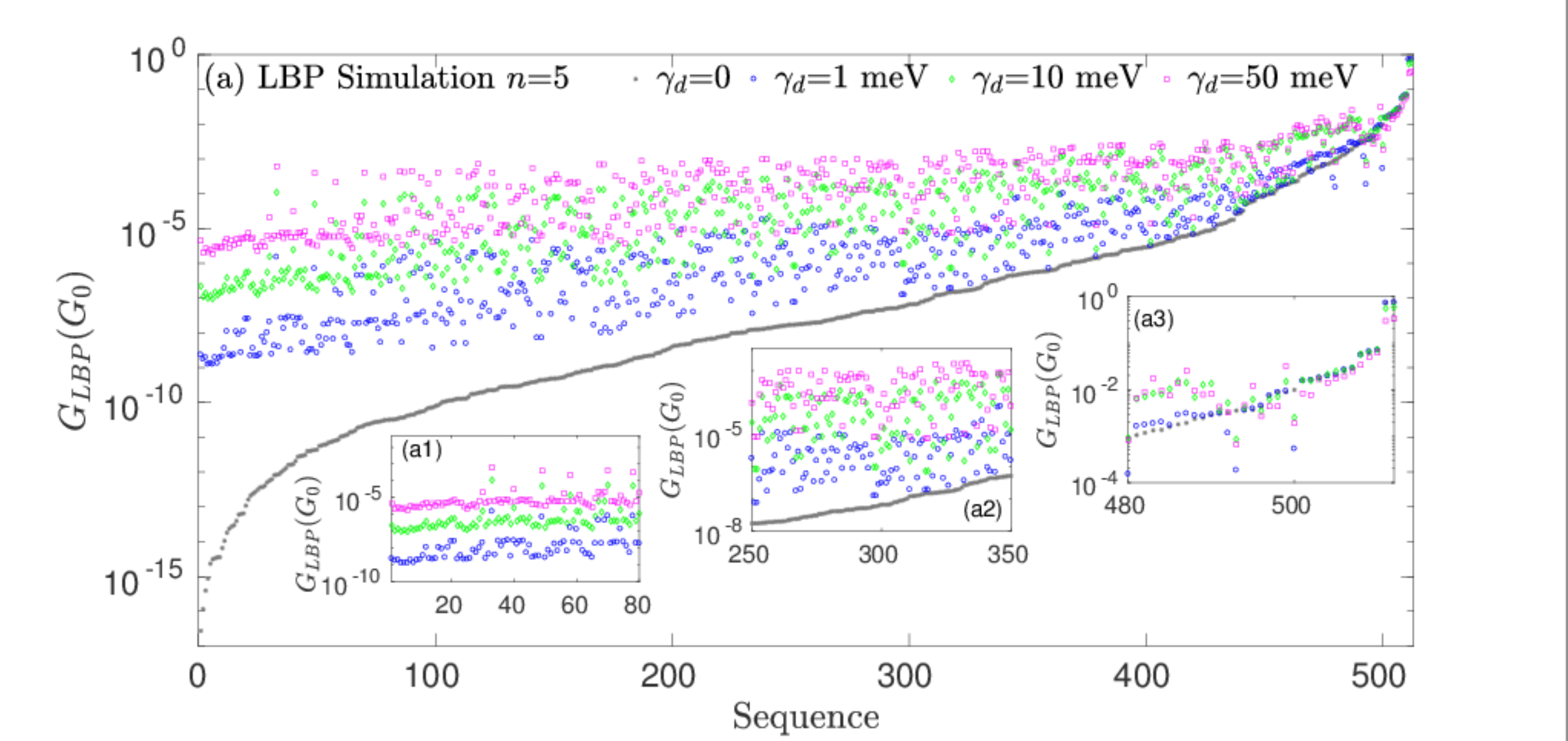}
\includegraphics[width=7cm]{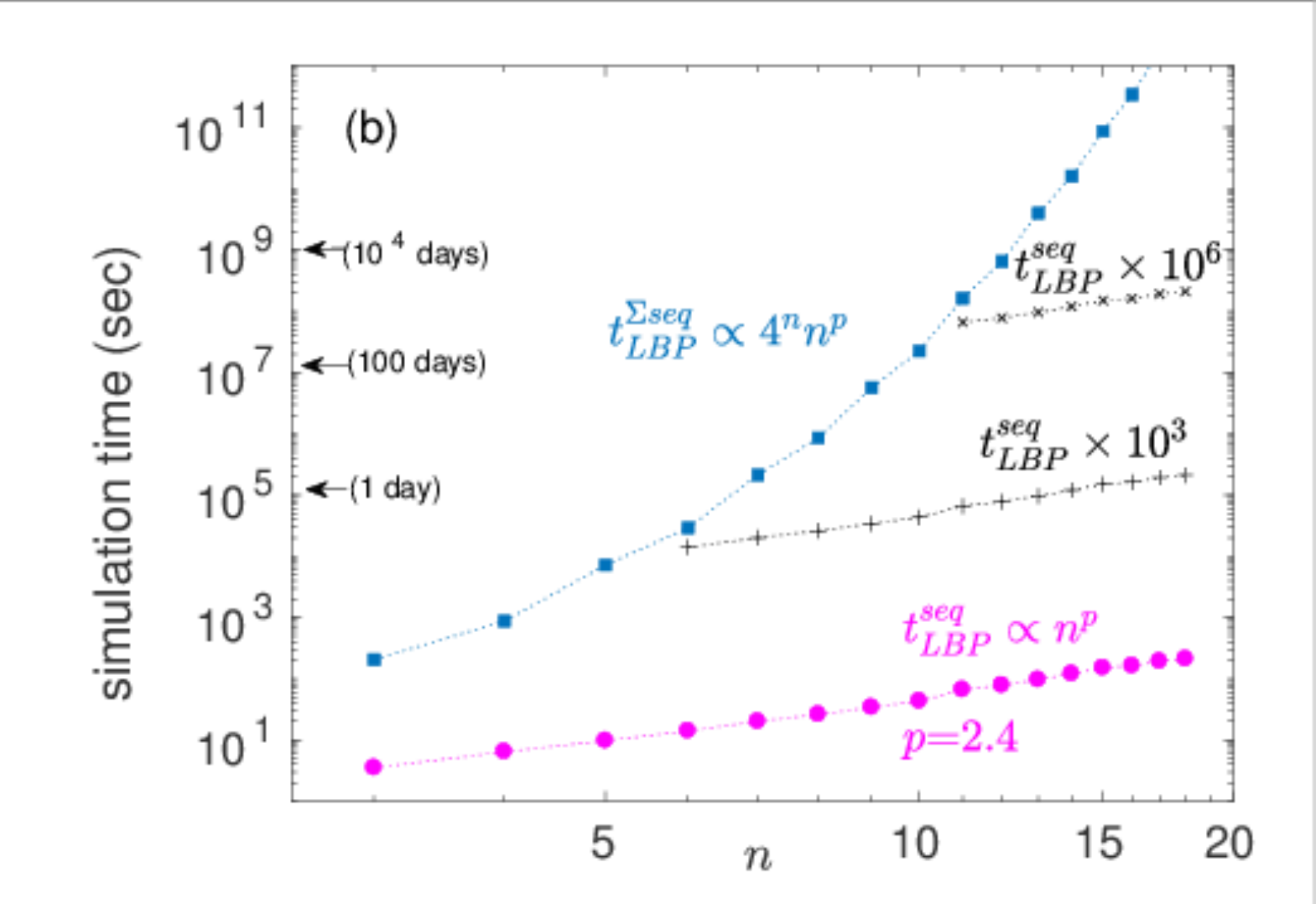}
\caption{LBP simulations.
(a) Effect of the thermal environment on the conductance of sequences with five base pairs
($n=5$, $N$=512).
Sequences are ordered by conductance values based on the case with environmental effects turned off.
(b)  Simulation times for a single sequence, $t_{LBP}^{seq}$ ($\circ$),
and for the whole ensemble of $N$ molecules with $n$ bp, $t_{LBP}^{\sum seq}$ ($\square$).
We further provide LBP simulation times for a subset of $10^3$ ($+$) and $10^6$ (x) sequences with $n$ bp. 
Simulation times are reported for $\gamma_d\neq0$. 
}
\label{FigLBP}
\end{figure*}

\subsection{Landauer B\"uttiker probe simulations} 

Charge migration in dsDNA takes place inside the double helix along the $\pi$ stacking. 
To preserve its structure thus conductive properties, dsDNA must be maintained in a buffered solution,
making it a highly flexible material with nuclear dynamics of the nucleobase, 
reorganization of solvent molecules and counterions, and displacements of structural elements. 
The timescales of these processes range from picoseconds to milliseconds. 

Simulations of charge carrier dynamics in DNA must therefore take into account 
the impact of atomic motion on charge conduction.
Ab-initio techniques are impractical to treat this problem, and even hybrid methods are challenging:
Explicit atomistic calculations at the level of molecular dynamics (for sampling configurations),
combined with quantum mechanics/molecular mechanics (QM/MM) methodologies 
are limited to short sequences and short evolution time \cite{cunibertiNJP10,kubar1,kubar2,kubar3}.
This challenge calls for the development of coarse grained models with perturbative methods
\cite{cunibertiphonon,ladder3,PeskinPCCP,Voityuk18},
model-system computational schemes \cite{kubar3}, and phenomenological approaches \cite{Kim1,Korol1,Korol3,Ryndyk18}.
For example, the fluctuating nuclear environment of DNA can be described by adding noise terms
to the static, purely electronic Hamiltonian. These fluctuations
can be made spatially and temporally correlated, 
mimicking coordinated atomic motion \cite{flickering,BeratanZhang}.

With the goal to describe charge transport in 1-5 nm long DNA, identify central transport
mechanisms, and pinpoint exceptionally good or poor DNA conductors,
we recently performed extensive conductance calculations of metal-molecule-metal DNA junctions \cite{Korol3}.
In our scheme, the electronic structure of the double helix is described with a coarse grained model \cite{BerlinJacs,Wolf},
and the electrical conduction is calculated with a quantum scattering technique, the Landauer-B\"uttiker's probe (LBP)
method \cite{Buttiker1,Buttiker2}. 
This approach captures the impact of intra and inter-molecular motion (``thermal environment")
by introducing a tunable scattering parameter, $\gamma_d$, responsible for
decohering the conducting charge and opening elastic and inelastic charge transport pathways
\cite{PastawskiD, PastawskiT,Dhar,Nozaki1,Qi13,Chen-Ratner,WaldeckF,Kilgour1,Kilgour2,
Kilgour3,Korol1,Kim2,Korol2,Kim1}. 
Physically, $\gamma_d$ corresponds to a scattering rate constant, 
but here we report it with the dimension of energy.

The LBP method as applied to molecular junctions was detailed in several previous 
publications from our group, starting in Refs. \cite{Kilgour1,Kilgour2,Kilgour3},
where general principles were explained. 
Applications to DNA junctions were discussed in Refs. \cite{Kim1,Kim2,Korol1}.
The code was presented in details in Ref. \cite{Korol2}.
LBP calculations performed in Ref. \cite{Korol3} comprises the training set for the present study.

Briefly, the electronic Hamiltonian of dsDNA is modeled using a tight-binding
ladder Hamiltonian with each site representing a particular base, see Fig. \ref{Fig1}(a).
The electronic energies of each site and the interstrand and intrastrand transition matrix elements $t$
were generated by DFT calculations \cite{BerlinJacs}.
%
The molecule-metal coupling strength $\gamma_{L,R}$ is taken in the range of  50-1000 meV,
and the electronic temperature $T_{el}$ at the metal leads is 5-300 K.
Information on thermal-environmental effects are all incorporated within the parameter $\gamma_d$,
taken in the range of 0-50 meV; the structure is rigid when $\gamma_d=0$.
Another tunable parameter is the position of the Fermi energy of the metals relative to
molecular states. Since the HOMO of the guanine nucleotide lies the closest to the Fermi energy of gold
electrodes, compared to the other nucleotides, we place the Fermi energy of the 
metals in resonance with the on-site energy of the guanine base.
The LBP procedure involves the solution of an algebraic equation.
It outputs the linear response (low voltage) electrical conductance $G_{LBP}$ of the junction.

Calculations reported in Ref. \cite{Korol3} were all-inclusive:
For a given  number of base pairs $n$, we considered all possible sequences
satisfying the base-pairing rules and simulated the electrical conductance of each of these compounds
using the LBP scheme. With this exhaustive approach, we discovered key 
principles governing 
charge transport in 3-7 bp DNA nanojunctions,  which are 1-3 nm long. 
As an example, in Fig. \ref{FigLBP} we display the conductance of dsDNA with $n=$5 bp under 
increasing environmental effects. There are $N=$512 such distinct sequences. As we discuss in Ref. \cite{Korol3},
we identify in this figure different families of molecules: good ($G=1-0.01$ $G_0$),
 poor ($G<10^{-10}$ $G_0$), and intermediate conductors ($G=10^{-2}-10^{-6}$ $G_0$). 
 Here, $G_0=e^2/h$ is the quantum of conductance per spin species,
 with the electron charge $e$ and Planck's constant $h$.
Good conductors (panel a3)
are only mildly affected by the environment showing a quantum-wire properties; poor conductors  (panel a1)
enjoy a dramatic enhancement of their conductance, assisted by 
the environment, demonstrating a tunneling-to-hopping crossover.
Nevertheless, the majority of the sequences cannot be 
categorized as tunneling/ohmic/ballistic conductors (panel a2),
and their conductance under environmental effects largely depends on whether they have a local
cluster of identical bases \cite{Korol3}.  
%

The cost of LBP calculations 
rapidly increases with molecular length 
as we illustrate  in Fig. \ref{FigLBP}(b).
Here, we report the LBP computation cost for a single sequence of length $n$, $t_{LBP}^{seq}$, as well as
the time it takes to generate $G_{LBP}$ for the whole ensemble of $N$ molecules with $n$ base pairs,
$t_{LBP}^{\Sigma seq}$.
Simulations were performed on a quad core processor, 
Intel(R) Core$^{ \rm { {\tiny TM}}}$ i5-6400 CPU \@ 2.70GHz.
For example, the calculation of $G_{LBP}$ for a
 single $n=7$ sequence takes 20 sec. 
 A complete scan of all 8,192 sequences with 7 bp takes about 2 days.
Continuing in this fashion, calculating $G_{LBP}$ for a single $n=12$ sequence
 takes 1.3 min , thus computing $10^6$ such values (a sub group of the full space)
 requires about 900 days. A full sequence scan for $n=12$ is impractical (see
supplementary material for more information).
All in all, it is clear that beyond $n=8$,
comprehensive calculations even with the highly simplified LBP method are impractical given
the exponential growth of the sequence space.
We next describe a ML approach that allows a rapid and accurate calculation of 
long range charge transport in DNA molecules.
\subsection{Machine learning approach}

Machine learning techniques are gaining much interest in the study of physical phenomena
and in the exploration of the chemical space.
For example, in condensed matter physics ML models detect,
classify, and characterize quantum phases and phase transitions in strongly correlated materials \cite{MLphase1,MLphase2}.
In quantum chemistry, machine learning tools construct potential energy surfaces,
identify reaction pathways \cite{MLreact,MLRoitberg,MLRoitberg18},
and predict excited state energies \cite{AAG16} and electronic correlation energies \cite{MLMiller}.
Application of ML methods to studies of materials  \cite{MLmat}
and drug design \cite{MLdrug}, e.g. predicting  drug-target binding affinities,
can automate molecular discovery and synthesis \cite{Doyle}.

Numerous applications of ML models are focused on materials with predictions made on the 
energy of the system: potential energy function, formation energy, electronic energy. In contrast,
the potential of ML models to explore chemical dynamics, specifically probe the question of structure-dynamic, 
is still largely untouched. Some examples include Ref.  \cite{MLQT}, where
a ML approach was used to predict the electrical conductance of disordered one-dimensional channels,
and Refs. \cite{AAG16,AAG17}, in which excitation energy dynamics and transfer times
in light harvesting systems were predicted with ML tools
by considering a large dataset of model Hamiltonians.

	
	
\subsubsection{Input representation: Descriptors for DNA charge transport}

Our objective is to predict the electrical conductances of DNA junctions of arbitrary length
based on the conductance of short DNA molecules of 3 to 7 bp.
The input data representation (descriptor)
should capture central features of DNA electrical transport.
We opt here for a descriptor that does not rely on 
an energy function of the system or its Hamiltonian, unlike e.g. Ref. \cite{AAG17}.
As such, the neural network (NN) can predict the conductance of long sequences---while trained on short chains.
Since the input features used here are missing the fingerprints of the methodology (no model parameters),
one could follow our steps, adopt the descriptors examined here, and train data generated from other methodologies 
besides the LBP method.

We devise and test different types of input features for DNA charge transport, which we identify as
single (S), pair (P) and  trio (T) descriptors.
The smallest descriptor has four numbers as its input, and it only
captures the composition of the sequence: We 
count the occurrences of each of the A,C,G, T nucleotides (in that order) in a single strand.
For example, the sequence 5'AAATGG3' has an input descriptor  $[3 0 2 1]$.
This descriptor is of type S, since it provides information on individual
base pairs, as opposed to the P and T input features, which hand over
information over local clusters.

The DNA ladder model involves nearest-neighbor intra- and inter-strand interactions.
Indeed, an important aspect of DNA charge transport is that the transferred
charge may be delocalized over several base pairs.
It is therefore expected that besides composition, information over the ubiquity of 
pairs and threesomes (e.g., the appearance of the series GGG in a sequence) 
are paramount to charge transport characteristics of DNA.
We therefore test P and T descriptors:
The occurrence of pairs of nucleotides is described by an $4^2$-dimensional vector;
the number of unique triplets is described by an $4^3$-dimensional vector. 
We list pairs and trios in alphabetical order: AAA, followed by AAC, AAG, AAT, ACA, etc. 
For example, the input descriptors S, P, and T of the sequence AAAAC are 
$[4 1 0 0]$, $[3 1 \v{0}_{1\times14}]$,
and $[2 1 \v{0}_{1\times62}]$, respectively.

So far, the descriptors do not convey information on the fact that the examined molecule is positioned in a
metal-molecule-metal configuration.
As we demonstrated in Ref. \cite{Korol3},  the identity of the nucleotide connecting 
to the electrodes is important, particularly for short chains under coherent transport.
Therefore, we expand the descriptors explained so far by
adding two parameters that specify the bases, A, C, G, or T,  connected to the electrodes. 

Altogether, we organize six models with $4-,6-,16-,18-,64-\text{ and }66-$ dimensional descriptors, 
which we refer to as the NN-4, NN-6, NN-16,..., NN-66 models. In the next section we test these models.
Generally, we observe that the NN-66 model provides the most accurate predictions.
	


\begin{figure}[htbp]
\includegraphics[width=9cm]{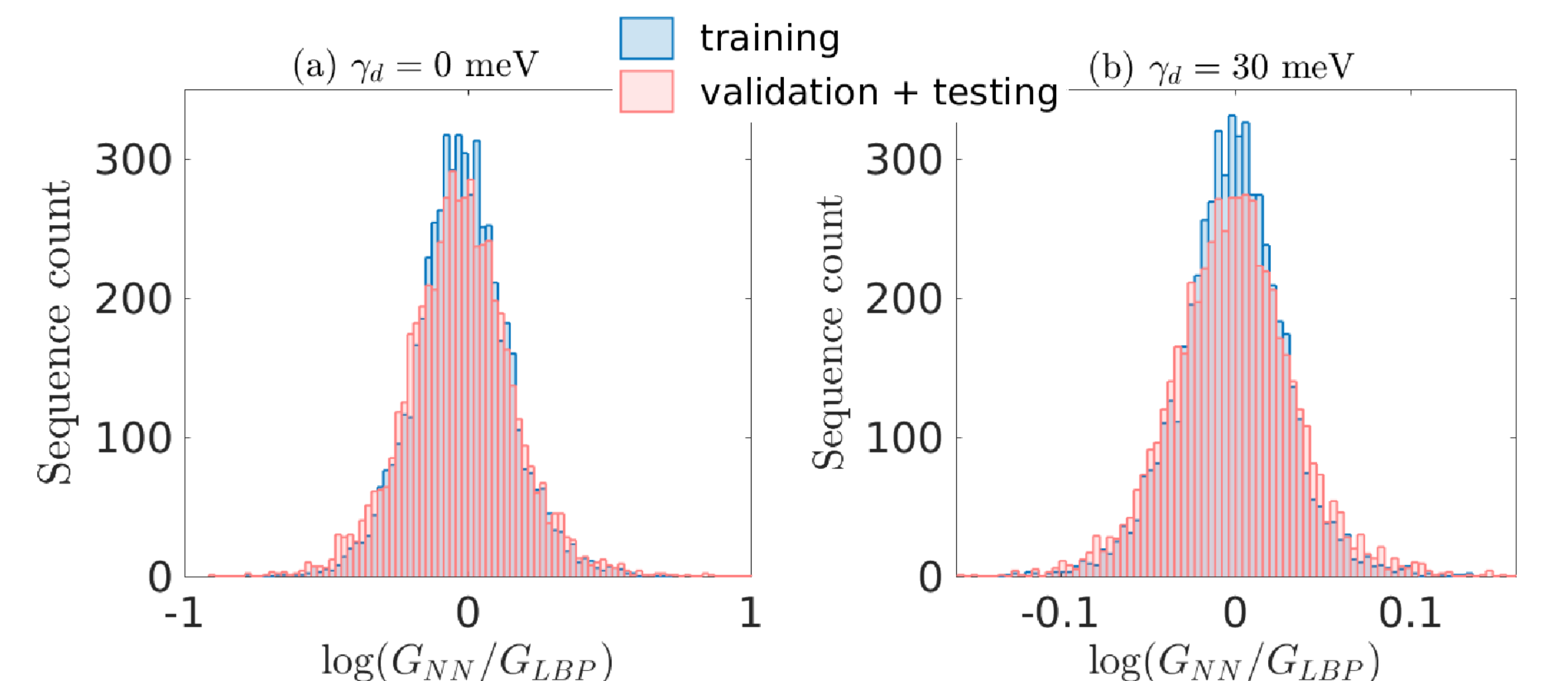}
\vspace{5mm}
\caption{Error distribution for training and validation+testing sets,
(a) $\gamma_d=0$, (b) $\gamma_d\neq0$.
Other parameters are $\gamma_{L,R}=50$ meV, $T_{el}=5$ K.
We trained 10 networks with the NN-66 descriptor.
}
\label{Fighist}
\end{figure}

\begin{figure*}[htbp]
\includegraphics[width=19cm]{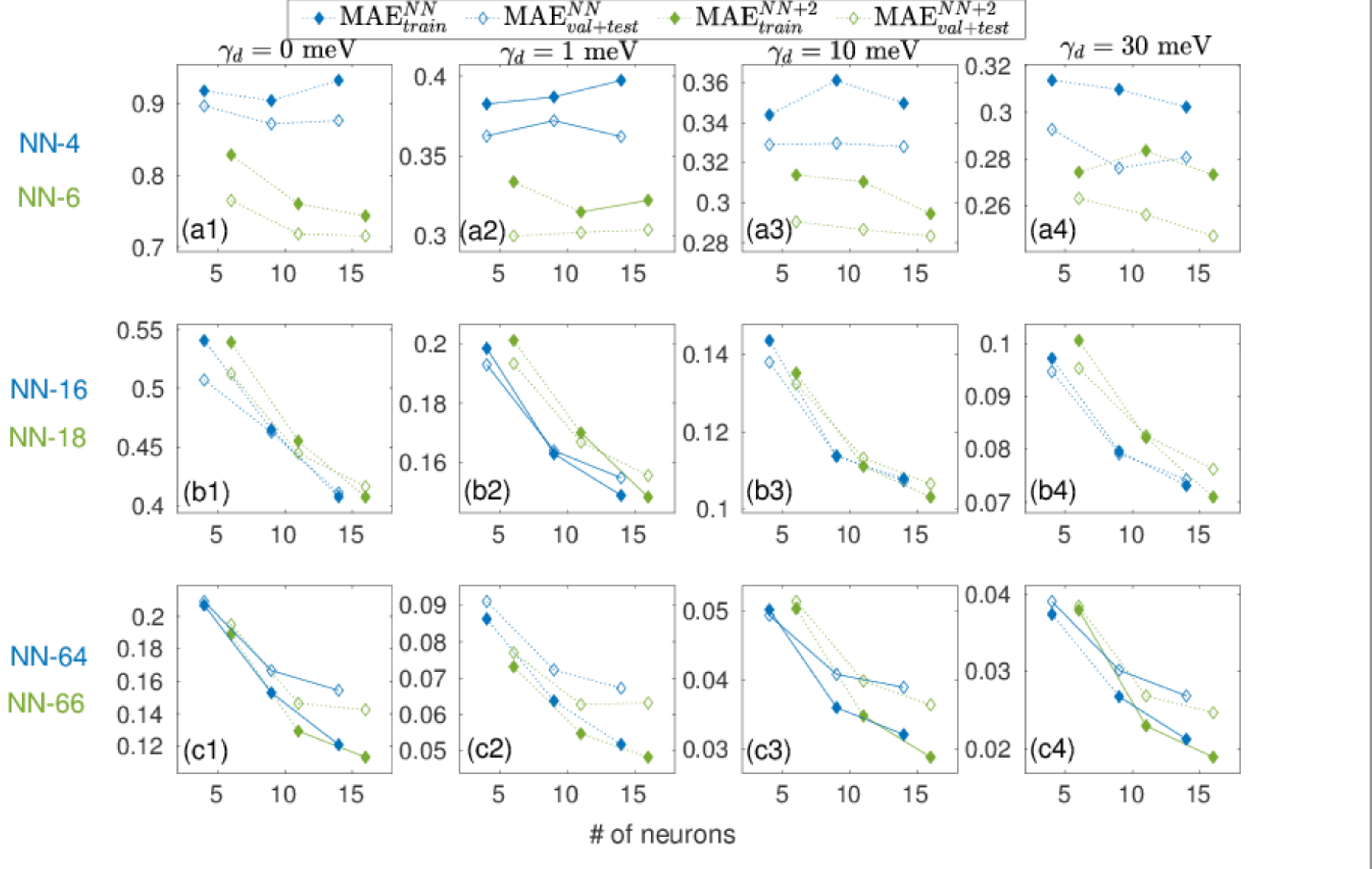}
 \caption{Performance of the NN  with 5, 10 and 15 neurons in its hidden layer (x-axis)
for different descriptors (rows).
The dataset includes all 10,952 sequences with $n=3-7$ bp  
under different environmental effects $\gamma_{d}$=0, 1, 10, 30  meV (columns).
We display the MAE  for both the
training set (50\% of the data) and validation+testing set (25\% each).
For clarity, results from different descriptors are shifted by $\pm 1$ about 5, 10, and 15.
Other parameters are $\gamma_{L,R}=50$ meV, $T_{el}$=5 K.
Corresponding $\sigma$ values are presented in the supplementary material file.
}
\label{Fig2}
\end{figure*}

\subsubsection{Neural network architecture}


We solve an input-output fitting problem with a shallow neural net consisting of a hidden
and an output layer.
The NN is trained using the Levenberg-Marquardt back propagation  
algorithm. 
The base-10 logarithm of the electrical conductance, obtained from the LBP method, is 
provided to the NN as the target value. 
 We used the built-in MATLAB machine learning toolbox ({\rm trainlm})
 with a random assignment of the LBP data:
50\% for training and 25\% for validation and testing each.
The algorithm adjusts the weights and biases from the initial guess to minimize the mean squared error,
that is the averaged square error between the network outputs
$a_i$ and the targets $t_i$ over a training set with $K$ values.
Since conductance  values (in $G_0$)  span up to 20 (!) orders of
magnitude, we use base-10 logarithm of conductance as our model's input and output.

To train the NN, we collect
 $10,952 = 32+136+512+2080+8192$ LBP calculations for DNA junctions with $n=3$ to $n=7$ bp.
The model is trained on these sequences with a random 2:1:1 assignment for training, validation and testing sets. 

We define deviations from the correct (LBP) value, $\Delta_i\equiv \log G_{NN}^{(i)}- \log G_{LBP}^{(i)}$.
This measure is Gaussian-distributed with mean $\approx0$. 
To characterize the model's accuracy and precision, we 
further define the mean absolute error (MAE) and the standard deviation (SD or $\sigma$) from the mean as
%
$MAE \equiv \langle |\Delta|\rangle$,
$\sigma^2 = \langle  \Delta ^2 \rangle$.
%
%
The mean absolute error discloses how many orders of magnitude away from the correct value is a typical prediction; 
we use the  modulus so as to capture absolute deviations (distance of values below and above) from the correct result. 
The standard deviation from the mean characterizes the width of the distribution (see e.g. Fig. \ref{Fighist}).

We train several NN models, for a specific choice of $\gamma_{L,R}$, $\gamma_{d}$ and $T_{el}$. 
Upon training, which takes under one second for the smallest descriptor and a little over 10 seconds for the largest one, 
the conductance values of millions of DNA sequences can be readily predicted, see Table \ref{table:n5}.

The weights and biases achieved upon training the NN may vary due to different initial guesses and the partition
of the data set into training, validation, and test sets. As a result, 
neural networks that are trained on the same data set give somewhat 
different predictions. 
We find that an interpolation task 
(when the NN predicts the conductance of sequences  of the same length as those trained on) 
is robust for our networks.
However, when asking the NN to extrapolate to out-of-sample, longer sequences, 
variations between predictions may be substantial,
by up to an order of magnitude when $\gamma_d$ is small (1 meV).
To address this issue, we train several (ten)  
neural networks on the same input data, and take the median of ten predictions as the model's prediction. 

In Fig.  \ref{Fighist}, we  display histograms of the training and validation-test errors $\Delta_i$.
Each prediction is the median of values
from 10 neural networks with 10 neurons in their hidden layer. 
The 10 neural networks are trained on the same data set,
but with different, random partitioning (2:1:1) into the training, validating and testing sets, respectively.
We confirm that deviations of the ML prediction from the LBP values are distributed 
symmetrically around zero. 
It is important to note that the error at nonzero 
$\gamma_d$ is significantly smaller than the rigid case, $\gamma_d=0$.
Similar results were obtained for other parameters.


In Fig. \ref{Fig2} we examine the performance of different descriptors, 
as well as the impact of the number of neurons in the hidden layer as
we vary $\gamma_d$ (columns).
We show the MAE; results for the standard deviation are included in the supplementary material file.
We examine six different descriptors.  In each case we also test whether information over the connectivity
(the base connected directly to the electrodes) is important for the quality of prediction.
For example, MAE=0.2 corresponds to $G_{NN}/G_{LBP}=10^{0.2}$, meaning that on average predictions are a
factor of 1.6 from the correct one. MAE=0.02 corresponds to $G_{NN}/G_{LBP}=10^{0.02}$, thus prediction are highly accurate with 
an average 5\% deviation from the correct value.
For clarity, the number of neurons used for the two different descriptors, NN and NN+2,
is displaced by $\pm 1$ from the actual numbers of 5, 10, 15.

We make the following observations:
(i) Scrolling down,  it is clear that the largest, 66-dimensional descriptor manifests the best performance.
This result is backed by physical knowledge over the role of small clusters of nucleotides in DNA charge transport.
(ii) The information over the connectivity to the metal is important, particularly 
when  $\gamma_d$ is small, as evidenced by the significant narrowing of the MAE and the SD for all three descriptor types. 
(iii) The error is reduced as we increase $\gamma_d$. 
The fully coherent case ($\gamma_{d}=0$) is the most challenging one for the ML model 
given the enormous range of conductance values.
(iv)
The number of neurons in the hidden layer is selected to ensure the best fit, but no over-fitting.
We set on the optimal number of 10 neurons. Larger networks do not substantially decrease the error 
for the validation and test sets. As one can see in Fig. \ref{Fig2} (for example panels b4 and c2), 
the performance of the 15-neuron networks is often not as good
for the validation+testing as it is for the training set, indicating over-training. 
On the other hand, 5-neuron nets are mostly under-trained. 

The ML framework could be improved in two ways. 
First, the training data was selected here at random from the full ensemble of molecules. 
However, the training set may be selected more carefully using e.g. principal component analysis as in Ref. \cite{AAG17}.
This would allow us to  (i) dismiss redundant information from sequences that are very similar in properties,
 and (ii) capture under-represented structures, such as homogeneous sequences.
Second, the ML algorithm parameters such as the activation function, number of layers, regularization parameters, were not optimized
here, besides the number of neurons in the hidden layer. Custumizing the NN architectures could enhance the quality of predictions.

\section{Results}
\label{Result}

\subsection{Interpolation and extrapolation predictions} 

We study the performance of the NN on different tasks. 
An interpolation prediction concerns training the NN on say 10,000 sequences of $n=10$ bp, which is a subset of the total  524,800 10-bp long sequences,
then asking it to predict the conductance of another $n=10$ sequence.
An extrapolation task, in contrast, concerns training the NN on sequences that are $n=3-7$ bp long, and using it to predict the conductance of longer molecules.
Obviously, an extrapolation task is more economic since generating the dataset for long chains is costly, see the supplementary information file.
We now show that our ML method can perform very well for both Interpolation and extrapolation tasks.

Beginning with an interpolation prediction,
we exemplify the performance of the ML method for short sequences ($n=5$) in Fig. \ref{FigNN5}.
These results should be compared to LBP calculations of Fig. \ref{FigLBP}.
We recall that LBP conductance of rigid molecules sets the ordering of sequences. Thus, the fact that
in Fig. \ref{FigNN5} the conductance of rigid molecules is not monotonic but it shows local fluctuations demonstrates
errors in prediction, which can be an order of magnitude away from the correct LBP value for poor conductors.
When environmental effects are included, predictions become quite accurate:
Compare panels (a1)-(a3) in Fig. \ref{FigNN5} to Fig. \ref{FigLBP}.
Specifically, the NN reproduces the tunneling-to-hopping behavior for sequences $0-80$, 
the quantum wire characteristics of good conductors (sequences $450+$), and the behavior 
of sequences in between, see panel (a2).

The advantage of the ML framework is made clear in Table \ref{table:n5}. 
Predicting the conductance of 10$^6$ sequences with $n=18$ bp takes several seconds (after training),
which is 7 orders of magnitude faster than direct LBP calculations, see Fig. \ref{FigLBP}(b).

\begin{widetext}
\begin{minipage}{\textwidth}
\begin{minipage}[b]{0.62\textwidth}
\hspace{-8mm}
\includegraphics[width=12cm]{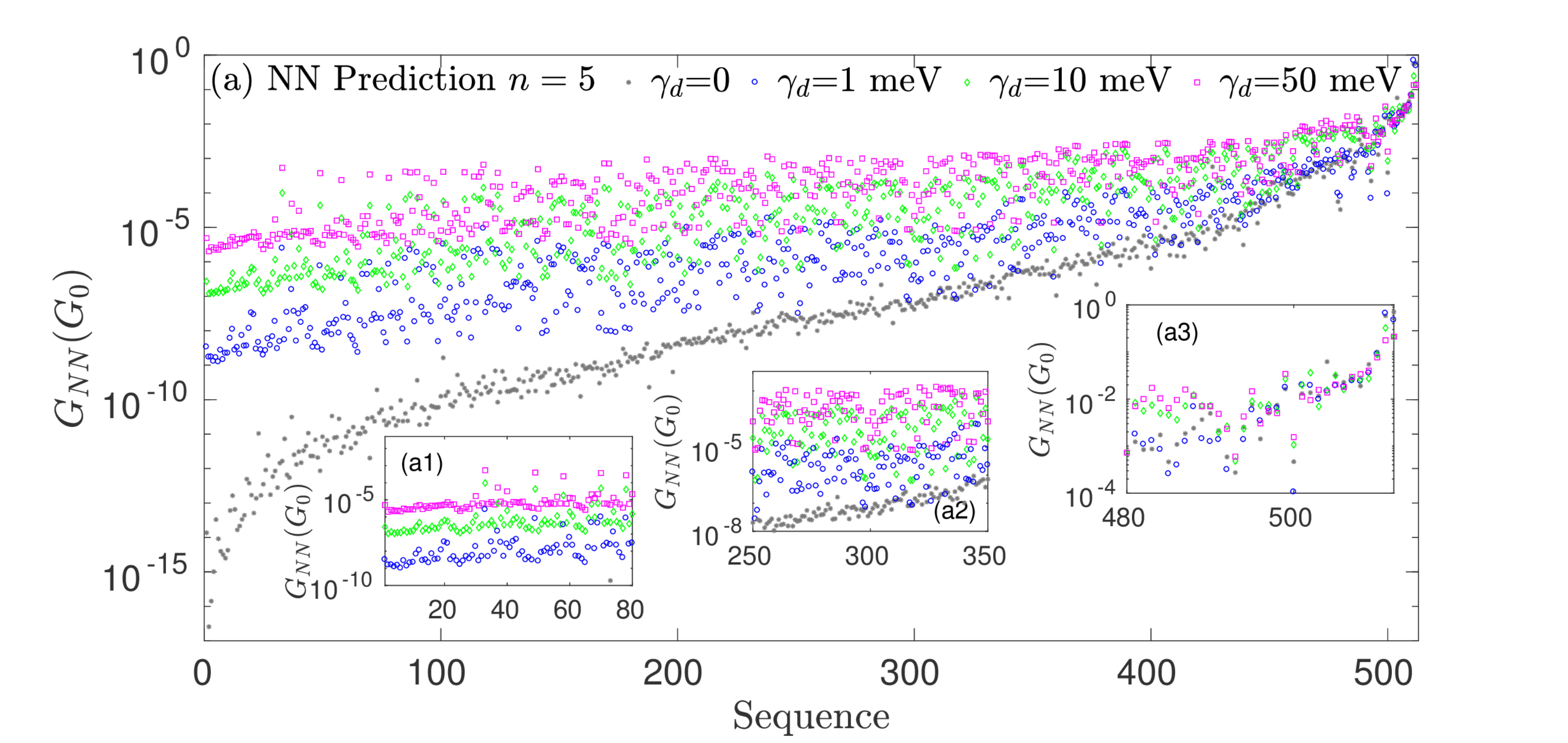}
    \captionof{figure}
%
{ML prediction. Effect of the environment on the conductance of sequences with five base pairs
($n=5$, $N$=512) as predicted by a
machine learning algorithm with 10 neural networks trained on the same 50\% of the data set,
but subdivided randomly (and differently for 10 networks) between training (70\%), validation (15\%) and test (15\%).
Sequences are ordered according to LBP simulations at $\gamma_d=0$.
}
\label{FigNN5}
\end{minipage}
 \hfill
\hspace{-10mm}
  \begin{minipage}[b]{0.32\textwidth}
    \centering
\begin{tabular}{|c| c| }
\hline
procedure  & time \\
\hline
building the dataset &  \\
$n=3-7$ (10,952 seq) &   \\
$\gamma_d \neq0$ &  2.3 days\\
$\gamma_d =0$ &  20 mins\\
\hline
training 10 NN & 2 min\\
\hline
prediction:  $10^6$  sequences &  \\
with 10 NN,  NN-6  & 1 sec\\
\hline
prediction:   $10^6$  sequences  &  \\
with 10 NN,  NN-66  & 10 sec \\
\hline
\end{tabular}
 \captionof{table}{{\bf 
Supervised training on a shallow NN using the
Levenberg-Marquardt back-propagation algorithm.  }}
 \label{table:n5}
 \end{minipage}
\end{minipage}
\end{widetext}

\begin{figure*}[htbp]
\vspace{2mm} \hspace{3mm}
\includegraphics[width=17cm]{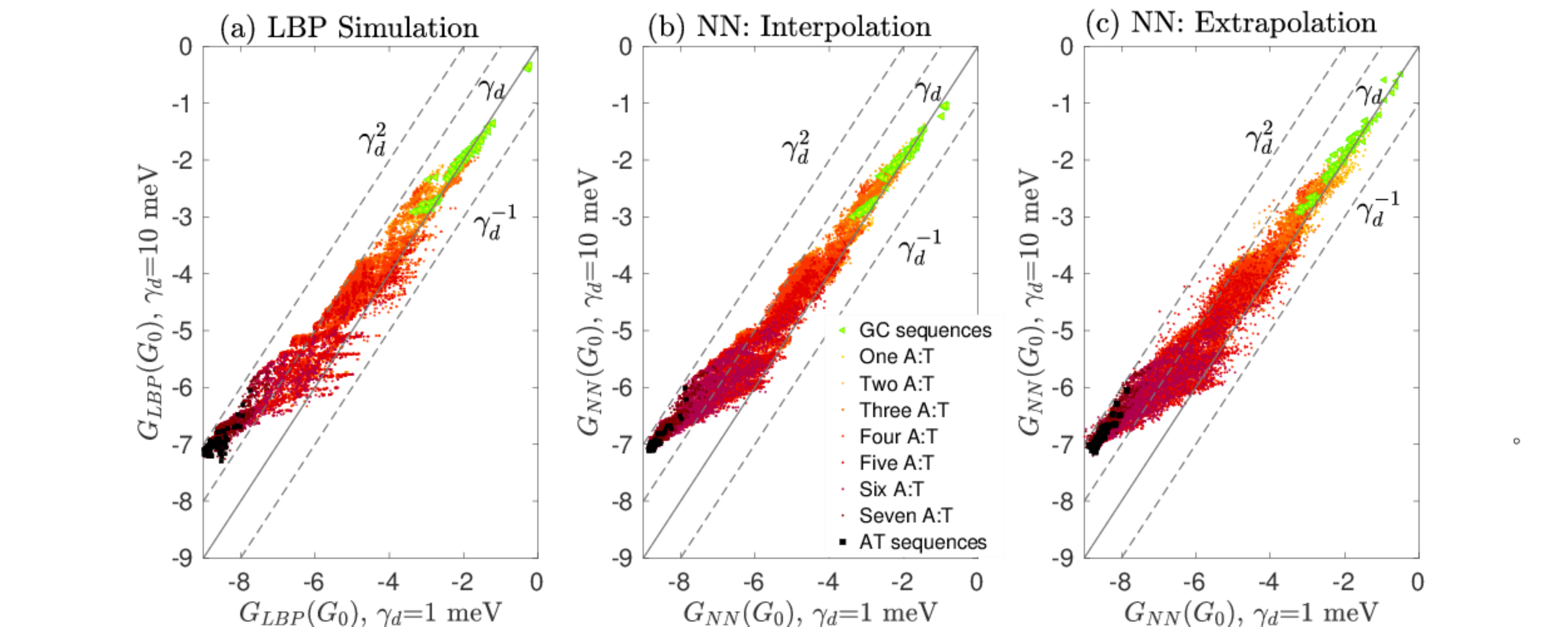}
\vspace{2mm}
\caption{Scaling of the conductance with environmental scattering rate $\gamma_d$ for sequences with eight base 
pairs ($n=8$, $N=32,896$),
(a) LBP calculations.
Machine learning framework, with NN trained on 10,952 sequences of (b) length 8 and (c) length 3-7. 
Other parameters are $T_{el}=300$ K, $\gamma_{L,R}=50$ meV.}
\label{FigNN8}
\end{figure*}

\begin{table*}[htbp]
 \begin{tabular}{|c|c|c|c|c|c|c|}
 \hline
 $\gamma_{L,R}$  (meV)  & $T_{el}$ (K) & $\gamma_d=0$ (meV) & $\gamma_d=1$ (meV) & $\gamma_d=10$ (meV) & $\gamma_d=30$ (meV) & $\gamma_d=50$ (meV) \\ 
   &  & MAE, $\sigma$ & MAE, $\sigma$ & MAE, $\sigma$ & MAE, $\sigma$ & MAE, $\sigma$ \\
 \hline 
 50  & 5  & 0.44,  0.63 & 0.17,  0.23 & 0.10,  0.14 & 0.071,  0.11 & 0.048,  0.075 \\ 
 \hline 
 50 & 300  & 0.71,  0.92 & 0.21,  0.30 & 0.085,  0.13 & 0.045, 0.081 & 0.028,  0.045 \\ 
 \hline 
 1000  & 5  & 0.50,  0.70 & 0.23,  0.31 & 0.16, 0.22 & 0.10, 0.14 & 0.080, 0.11 \\ 
 \hline 
 1000 & 300  & 0.70,  0.90 & 0.28,  0.38 & 0.13,  0.19 & 0.076,  0.12 & 0.56,  0.093 \\  
 \hline		
\end{tabular}
\caption{Mean absolute error and standard deviation of the mean error
for extrapolation predictions $G_{NN}$ for 8 bp long sequences.}
\label{table8}
\end{table*}


The question of an extrapolation prediction (for out-of-set $n$), as opposed to
an interpolation is examined in Fig.
\ref{FigNN8}. Here, we display the conductance of $n=8$ bp system (total of $N=32,896$ sequences)
for two different values of $\gamma_d$, 1 and 10 meV. Each dot corresponds to a particular sequence.
We begin in panel (a) with an all-inclusive LBP simulation.
The best conductors are found on the diagonal, or slightly below it;
they are either undisturbed or lightly negatively affected by thermal-incoherent scattering effects.
Poor conductors follow the scaling $G\propto \gamma_d^2$.
Most importantly, as reported in Ref. \cite{Korol3},
while we can identify ballistic (green) and ohmic (black) conductors, most molecules
display an in-between behavior, $G\propto \gamma_d^{\alpha}$ with $0\lesssim\alpha\lesssim1$
at high electronic temperature.
These sequences conduct via an {\it intermediate}, coherent-incoherent mechanism.

We use the ML model to reproduce these results.
In panel (b), the dataset was generated for $n=8$ bp molecules,
and it includes the conductances of 10,952 sequences (out of the total of  32,896).
In contrast, in panel (c) we train the NN on 10,952 short sequences with $n=3-7$ bp.
The three panels therefore present physical results, ML-interpolation,  and ML-extrapolation predictions.

Overall, we find that both predictions are qualitatively correct.
The ML algorithm performs very well for most of the sequences that display intermediate coherent-incoherent conduction.
It also successfully identifies poor hopping and good ballistic conductors.
Table  \ref{table8} lists errors associated with the out-of-set (extrapolation) predictions.

\subsection{Transport mechanisms}

We use the NN model to extrapolate and predict conductance trends in families of 
DNA molecules with $n=3-18$ base pairs. We emphasize that training is done on short sequences
with $n=3-7$ base pairs. The ML model is therefore expected to generalize results to out-of-sample sequences.
We focus on three families of molecules:

(i) Sequences with an AT block, which display a tunneling-to-hopping crossover with increasing barrier 
length, as demonstrated experimentally in Ref. \cite{Tao16} and computationally (with the LBP method)
in Ref. \cite{Korol1}.
Our results are summarized in Fig. \ref{Fig7} with Table \ref{table:seqAT} listing examined sequences.

(ii) Alternating GC sequences, which display ohmic resistance. 
The electrical conductance of this family of molecules was recently measured in
Ref. \cite{Tao16}, with calculations reported in Refs. \cite{Kim1,Korol1}.
Our simulations of this family are summarized in Fig. \ref{Fig6} with Table \ref{table:seqGC} listing the molecules.
%

(iii) Sequences with a uniform A block. 
The conductance of these homogeneous junctions was examined in Ref. \cite{Kim2}, showing complex behavior:
When environmental effects are weak ($\gamma_d\sim 0.001$ eV), we watch
a transition from deep-tunneling to resonant tunneling motion with increasing length.
In contrast, when $\gamma_d=10-50$ meV,  
we observe a crossover from tunneling to hopping conduction, similarly to case (i).
Our results are summarized in Fig. \ref{Fig8} with Table \ref{table:seqAA} listing the molecules.

We now discuss our observations.
Fig. \ref{Fig7} exemplifies the excellent ability of the NN model to predict transport mechanisms
and provide accurate predictions for the conductance.
It is significant to note that the conductance varies over 5 orders of magnitude. 
In panel (b) we display the ohmic behavior taking place when $\gamma_d=30-50$ meV. 
The NN-66 model generally provides more accurate predictions that the NN-18 model, 
even when $\gamma_d$ is large and charge transport is expected to be less delocalized over multiple sites.

Chains with alternating GC sequences are studied in Fig. \ref{Fig6}.
Experiments demonstrated that this system manifests 
an ohmic behavior with the resistance scaling as $R\propto n$, in accord with a site-to-site
hopping conduction mechanism \cite{Tao16}.
LBP calculations quantitatively reproduce these observations  \cite{Kim1,Korol1}.
We compare ML predictions to LBP simulations, showing mixed results.
The NN model predicts an approximate linear enhancement of 
resistance with length when $\gamma_d=10-30$ meV, but it misses this trend for smaller or larger values of $\gamma_d$. 

Why are ML predictions here less accurate than in Fig. \ref{Fig6}?
First, one should appreciate that results of the ML model are mostly within $\pm 2 M\Omega$ for
$\gamma_d=10-30$ meV.
We can rationalize the reduced accuracy of the NN model as follows. 
Recall that the training set includes all different combinations of the four bases for sequences 3-7 bps.
This ensemble of molecules realizes
conductances ranging from $1 G_0$ to $10^{-7} G_0 $ for e.g. $\gamma_d=10$ meV.
In Fig. \ref{Fig6},  however, we study a very specific subset of this ensemble, focusing 
on sequences that comprise a cluster of GC bps, with conductances that only mildly vary with distance. 
Predicting the behavior of a small, specific subset of molecules is therefore not trivial since they are not well represented
in the training set.
In principle, one could train the NN separately only on (short) GC molecules, to predict the behavior of long GC systems.
However, for $n=3-7$ there are only $\sim 100$ GC molecules, which is insufficient for a proper training and prediction.
More practically, the training set could be carefully prepared so as to ensure its diversity. 
All in all, we conclude that predicting the behavior of a specific
subset of molecules out of the full ensemble is challenging, but that the NN model
performs reasonable well for $\gamma_d=10-30$ meV.

In Fig. \ref{Fig8}, we study the  5'-G(A)$_n$G-3' family with $n=3-14$.
Here, the bridge is uniform, thus it can support a resonant tunneling (band like) motion.
This system displays three different transport regimes as we discussed in Ref. \cite{Kim2}: 
(i) For short junctions, tunneling conduction dominates with a strong suppression 
of conductance with length. (ii) As long as the environment only lightly influences the system,
$\gamma_d\lesssim 1$ meV, a resonant tunneling mechanism takes over deep tunneling in long enough chains.
For a uniform bridge, resonant tunneling manifests itself with a distance independent conductance.
(iii) In contrast, when $\gamma_d$ is large, charge carriers hop from site to site, showing an ohmic trend for large $n$. 
Experimentally, it was recently demonstrated that the conductance of 
adenine-stacked RNA-DNA hybrids varied weakly with length \cite{HihathA}, in what was termed as
`coherence-corrected hopping' mechanism \cite{Ratner15}.

We find that both NN-18 and NN-66 are quite successful in providing qualitatively (and even quantitatively) 
correct predictions. 
For short, $n<5$, chains, the NN-18 input descriptor 
better performs than the NN-66, which is not surprising given that for such short systems a triplet information is 
redundant as most input entries are null. In contrast, for $n>7$ the NN-66 is typically more successful than
the pairwise model.
However, in terms of transport mechanisms, similarly to Fig. \ref{Fig6},
the ML model again is missing the ohmic characteristics: It captures the deep tunneling trend for short systems 
and the ballistic saturation for $n>5$. However, it is not able to produce the ohmic trend for $\gamma_d\gtrsim 30$ meV.

We conclude that the NN model performed excellently in case (i)
when molecules were composed quite evenly of the different bases, but its predictions were less accurate 
in the other two cases, when the molecules were predominantly composed of either GC or AT bp. However, even 
when missing the correct distance dependence, predictions were generally accurate (correct order of magnitude)
for the NN-66 model. Ensuring that the training set includes unique, underrepresented sequences should improve predicitons.

%
\begin{widetext}

\begin{minipage}{\textwidth}
\begin{minipage}[b]{0.7\textwidth}
\hspace{-8mm}
\includegraphics[width=12cm]{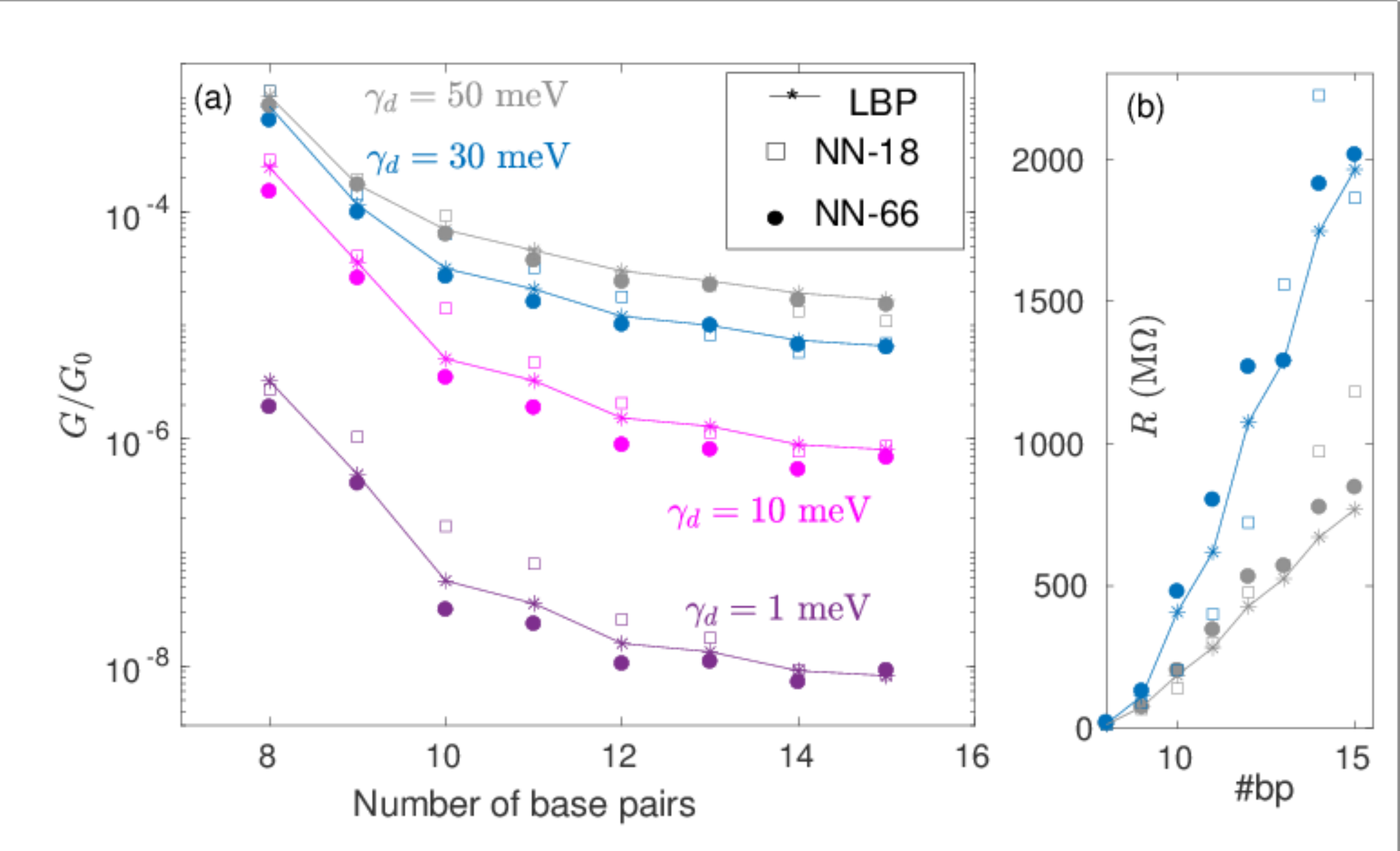}
    \captionof{figure}
{Tunneling-to-hopping crossover in DNA junctions.
(a) Conductance (log scale) of AT-block sequences (see Table \ref{table:seqAT})
as a function of the number of base pairs (\#bp). 
(b) Highlighting the ohmic resistance at high $\gamma_d$.
We compare LBP simulations ($*$) to NN-18 ($\square$) and NN-66 ($\circ$) predictions.
Simulations were performed at $T_{el}$=5 K to attenuate the contribution of ballistic electrons.
Other parameters are  $\gamma_{d}$=1, 10, 30, 50 meV, $\gamma_{L,R}$=50 meV.
}
\label{Fig7}
\end{minipage}
 \hfill
  \begin{minipage}[b]{0.25\textwidth}
    \centering
\begin{tabular}{|c| c| }
	\hline
	  $\#$ bp   &  5'-seq-3' \\ 
	\hline
	8 &  ACGCAGCGT \\
	\hline
	9&  ACGCATGCGT\\
	\hline
	10 &    ACGCATAGCGT \\
	\hline
	11 &    ACGC(AT)$_2$GCGT \\
	\hline
	12 &    ACGC(AT)$_2$AGCGT \\
	\hline
	13 &    ACGC(AT)$_3$GCGT \\
	\hline
	14 &    ACGC(AT)$_3$AGCGT \\
	\hline
	15 &   ACGC(AT)$_4$GCGT \\
	\hline
\end{tabular}
 \captionof{table}{{\bf AT-block sequences (seq) with 8-16 base pairs measured in Ref. \cite{Tao16}. 
}}
\label{table:seqAT}
    \end{minipage}
  \end{minipage}

%
\vspace{10mm}
\begin{minipage}{\textwidth}
\begin{minipage}[b]{0.7\textwidth} 
\hspace{-8mm}
\includegraphics[width=13cm]{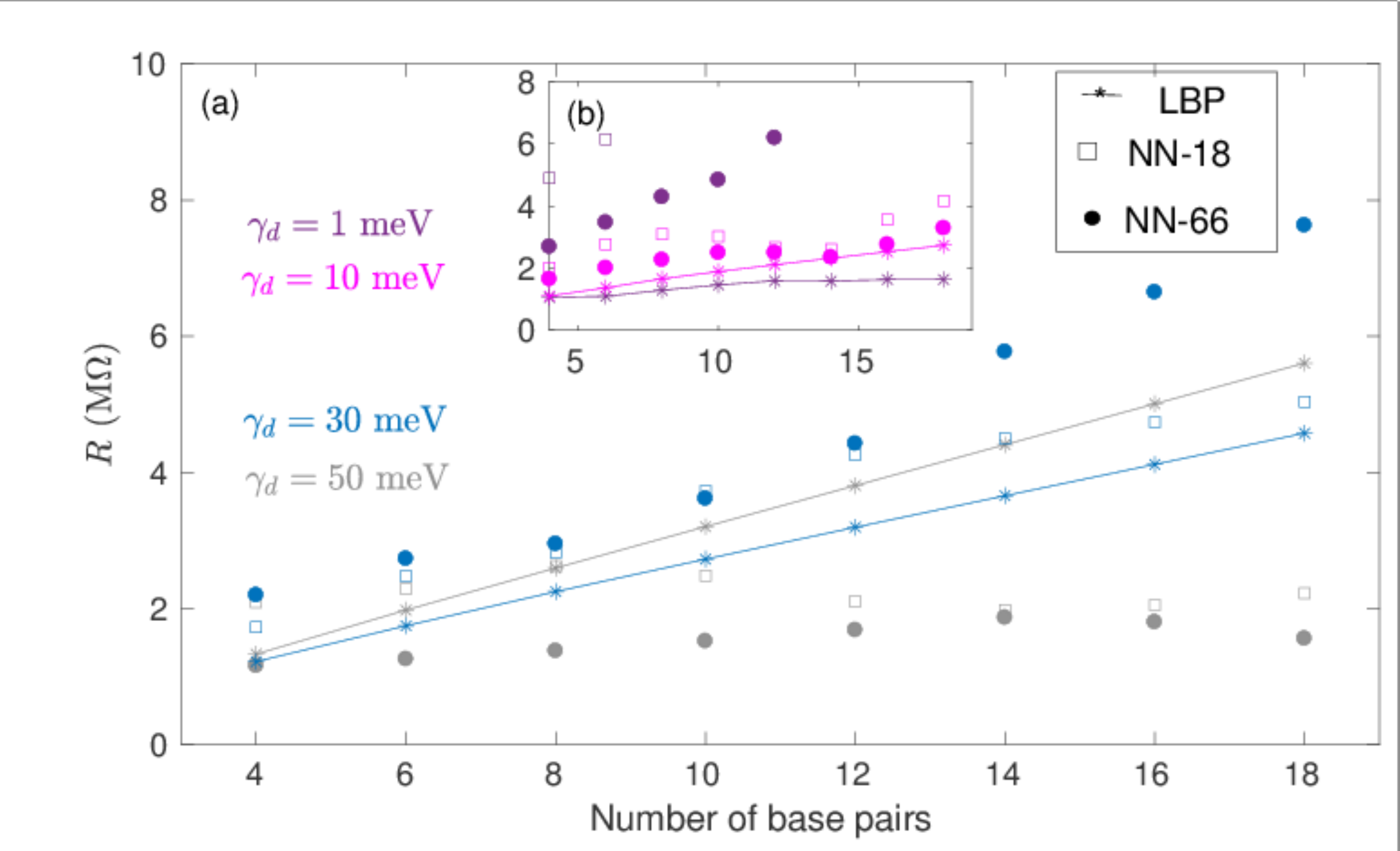}
   \captionof{figure}
{(a)-(b) Electrical resistance 
of alternating GC-DNA sequences as a function of length, 5'-A(CG)$_m$T-3' with $m=1-8$.
LBP simulations ($*$), NN-18 ($\square$) and NN-66 ($\circ$) models.
Other parameters are  $\gamma_{d}$=1, 10, 30, 50 meV, $T_{el}=300$ K,
$\gamma_{L,R}$=50 meV.
}
\label{Fig6}
\end{minipage}
 \hfill
  \begin{minipage}[b]{0.25\textwidth}
    \centering
\begin{tabular}{|c |c|}
        \hline
        $\#$ bp    &  5'-A(CG)$_m$T-3' \\
        \hline
        4  &  ACGT \\
        \hline
        6  & A(CG)$_2$T \\
        \hline
        8  & A(CG)$_3$T \\
        \hline
        10  & A(CG)$_4$T \\
        \hline
        12  &  A(CG)$_5$T \\
        \hline
        14  & A(CG)$_6$T \\
        \hline
        16  &   A(CG)$_7$T \\
        \hline
        18  & A(CG)$_8$T \\
        \hline
\end{tabular}
\captionof{table}{{\bf   Alternating GC-sequences measured in Ref. \cite{Tao16} with calculations reported in Fig. \ref{Fig6}.}}
\label{table:seqGC}
    \end{minipage}
  \end{minipage}

\vspace{5mm}
\begin{minipage}{\textwidth}
 \begin{minipage}[b]{0.7\textwidth} 
\hspace{-8mm}
\includegraphics[width=13cm]{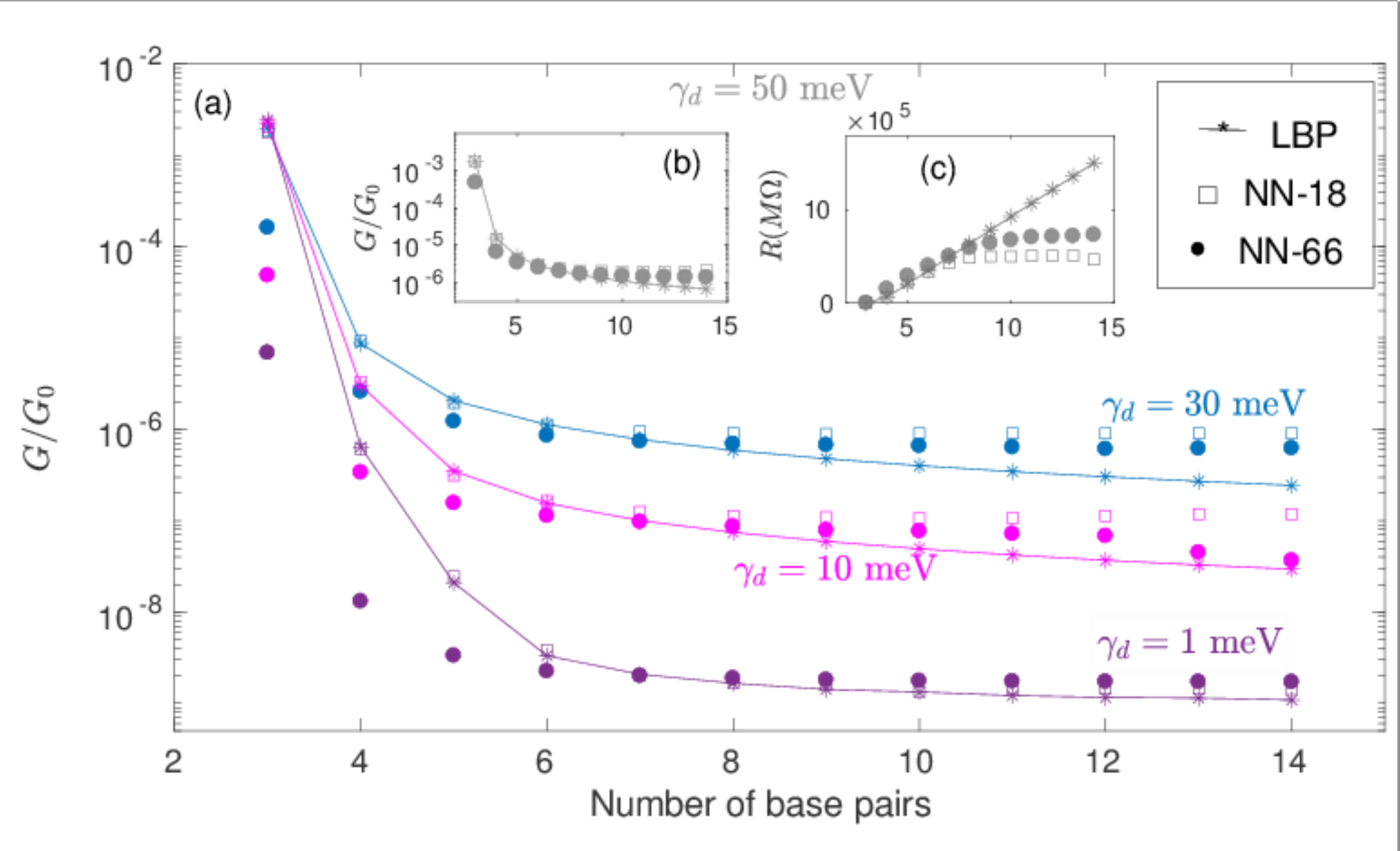}
   \captionof{figure}
{Electrical conductance of sequences with an A block, 5'-G(A)$_m$G-3', 
$m=1-12$.
as a function of number of bp.
We compare the NN-18 ($\square$) and NN-66 ($\circ$) predictions to  LBP simulations ($*$).
(b)-(c) Focusing on $\gamma_d=50$ meV, we demonstrate that the NN models fail to capture the ohmic behavior, 
but yield results within the correct order of magnitude. 
Other parameters are  $\gamma_{d}$=1, 10, 30, 50 meV, $T_{el}$=300 K, $\gamma_{L,R}$=50 meV.
}
\label{Fig8}
%
\end{minipage}
 \hfill
  \begin{minipage}[b]{0.25\textwidth}
    \centering
\begin{tabular}{|c| c| }
	\hline
	$\#$bp &  5'-G(A)$_n$G-3' \\
&   $n=1-12$ \\ 
	\hline
	3  &  GAG \\
	\hline
	4  &  G(A)$_2$G \\
	\hline
	5  & G(A)$_3$G \\
	\hline
	6  &  G(A)$_4$G \\
	\hline
	7  & G(A)$_5$G \\
	\hline
        ... & ...\\
        \hline
	 14  & G(A)$_{12}$G \\
	\hline
\end{tabular}
\captionof{table}{{\bf Sequences with a uniform A block, studied previously
 in Ref. \cite{Kim2} with the LBP method, and examined with the NN models in Fig. \ref{Fig8}.}}
 \label{table:seqAA}
    \end{minipage}
  \end{minipage}
\end{widetext}

\section{Summary}
\label{Summ}

We showed that a NN model could
provide cheap predictions of molecular conductance of millions of long DNA nanojunctions with high accuracy.
A central aspect of our work has been to test
input features (descriptors) that do not rely on model parameters (such as the coarse grained Hamiltonian)
and do not rightly announce on the base sequence. 
The developed descriptors provide
information over molecular composition and the occurrence of
local (2-3 bp) clusters.
As such, the model is transferable to sequences of different length and to data generated from other microscopic methods.


The ML model was trained on the conductance of 3 to 7 bp sequences. It was tested
against short molecules in this set, then used for predicting the
behavior of $n=8$ molecules. 
Moreover, we studied DNA sequences with 3-18 bp and
demonstrated that the ML model could well reproduce the behavior of special sequence subsets:
It traced the transition from fully coherent (deep tunneling) to ballistic (on-resonance) transfer
for stacked A-sequences \cite{Kim2}, and the tunneling to hopping crossover in sequences with an
AT block \cite{Korol1}. With mixed success, it recovered the 
hopping conduction in alternating GC-sequences \cite{Kim1}.
These results demonstrate that a ML method can provide a 
cheap assessment of quantum transport in DNA junction. 

It would be interesting to generate datasets using other (relatively cheap) methods that 
perturbatively include electron-vibration interaction effects, see e.g. \cite{Gauger18,Peskin1,Peskin2}
and explore the ability of the ML  technique to reproduce other transport mechanisms.

Before concluding, its worth recalling 
other types of measurements of charge migration in DNA besides the junction geometry, using 
electrochemistry \cite{BartonRev,WaldeckRev} 
or time-resolved spectroscopy \cite{Lewis08,Lewis}. 
In analogy to our study, one could perform quantum dynamics and charge transfer calculations in short molecules using
e.g. a quantum master equation approach, then generalize results to other sequences  with an ML framework.
Finally, the principles outlined in this work could be applied to other types of oligomers, biomolecules in particular,
in an effort to explain and predict long-range, macroscopic electron transport behavior \cite{Allon,Amdursky}.

\begin{acknowledgments}
D.S. acknowledges support from the Natural Sciences and Engineering Research Council Canada (NSERC) 
Discovery Grant and the Canada Research Chair program.
The work of R.K. was supported by the Centre for Quantum Information and Quantum Control (CQIQC) summer fellowship at
the University of Toronto and the University of Toronto Excellence Research Fund. \\
\end{acknowledgments}

	
{}

\end{document}